\documentclass[fleqn,usenatbib,useAMS,letter,onecolumn]{mnras}

\usepackage{graphicx}	
\usepackage{amsmath}	
\usepackage{amssymb}	
\usepackage{multicol}        
\usepackage{bm}		
\usepackage{pdflscape}	

\def\planck{\textit{Planck}}

\newcommand{\vn}{\hat{\mathbf{n}}}
\newcommand{\vvel}{\mathbf{v}}
\newcommand{\vk}{\mathbf{k}}
\newcommand{\vrv}{\mathbf{r}}
\newcommand{\dz}{\delta z (\vn)}
\newcommand{\Fcal}{{\cal F}}

\def\apj{{ApJ}}                 
\def\apjl{{ApJ}}                
\def\aap{{A\&A}}                

\def\jcap{{J. Cosmology Astropart. Phys.}}
\def\mnras{MNRAS}             
\def\prd{{Phys.~Rev.~D}}        
\def\nat{{Nature}}              

\usepackage[T1]{fontenc}
\usepackage{ae,aecompl}

\usepackage{newtxtext,newtxmath}

\title[Density Weighted Angular Redshift Fluctuations]{Density Weighted Angular Redshift Fluctuations: a New Cosmological Observable.}

\author[C.Hern\'andez-Monteagudo et al.]{Carlos~Hern\'andez--Monteagudo$^{1,2,3}$
\thanks{Contact e-mail address: \href{mailto:chm@iac.es}{chm@iac.es}},
Jon\'as Chaves-Montero$^{4}$,
and Ra\'ul E. Angulo$^{5,6}$.
\\
$^{1}$Centro de Estudios de F\'\i sica del Cosmos de Arag\'on (CEFCA), Unidad Asociada al CSIC, Plaza San Juan, 1, planta 2, E-44001, Teruel, Spain\\
$^{2}$Instituto de Astrof\'\i sica de Canarias, La Laguna, E-38205, Tenerife, Spain\\
$^{3}$Departamento de Astro\'\i sica, Universidad de La Laguna, E-38206, Tenerife, Spain\\
$^{4}$HEP Division, Argonne National Laboratory, 9700 South Cass Avenue, Lemont, IL 60439, USA\\
$^{5}$Donostia International Physics Centre (DIPC), Paseo Manuel de Lardizabal 4, 20018 Donostia-San Sebastian, Spain.\\
$^{6}$IKERBASQUE, Basque Foundation for Science, E-48013 Bilbao, Spain.}
\date{\today}

\pubyear{2020}

\begin{document}
\label{firstpage}
\pagerange{\pageref{firstpage}--\pageref{lastpage}}
\maketitle

\begin{abstract}
We propose the use of angular fluctuations in the galaxy redshift field as a new way to extract cosmological information in the Universe. This new probe $\dz $consists on the statistics of sky maps built by projecting redshifts under a Gaussian window of width $\sigma_z$ centred upon a redshift $z_{\rm obs}$, and weighted by the galaxy density field.
We compute the angular power spectrum of the $\dz$ field in both numerical simulations and in linear perturbation theory. From these we find that the $\dz$ field: {\it (i)} is sensitive to the underlying density and peculiar velocity fields; {\it (ii)} is highly correlated, at the $\gtrsim 60\,\%$ level, to the line-of-sight projected peculiar velocity field; {\it (iii)} for narrow windows $(\sigma_z < 0.03$), it is almost completely uncorrelated to the projected galaxy angular density field under the same redshift window; and {\it (iv)} it is largely unaffected by multiplicative and additive systematic errors on the observed number of galaxies that are redshift-independent over $\sim\sigma_z$. We conclude that $\dz$ is a simple and robust tomographic measure of the cosmic density and velocity fields, complementary to angular clustering, that will contribute to  more complete exploitations of current and upcoming galaxy redshift surveys.
\end{abstract}




{\it {\bf Introduction.}} Redshifts in the electromagnetic spectrum of cosmic objects has been essential for the development of modern cosmology. During the third decade of the last century, E. Hubble found that redshifts in the spectra of galaxies ({\it nebulae}) correlated with their distance \cite{hubble_1929}. This supported Lema\^itre's suggestion of an expanding universe \cite{lemaitre27}, which had been motivated by Friedmann's dynamical cosmological models \cite{friedmann22} framed in the (by that time recent) formulation of Einstein's general theory of relativity.
This linear relation between the redshift of galaxies and their distance to the observer was interpreted as a clear piece of evidence for the expansion of the Universe. Galaxy redshifts were then seen as signposts of the {\it Hubble flow}, and since then they have been used as  distances estimators in the cosmos when interpreted within the {\it fireball} model of the universe, in which space has been expanding ever since an initial stage of arbitrary high density and temperature \cite{lemaitre31}. In the last quarter of the last century, the redshift induced by the peculiar motion of galaxies, triggered by the local gravitational field, was added to the Hubble flow redshift when describing the spatial clustering of luminous matter in the universe \cite{kaiser1987}. These peculiar motions induced the so-called ``redshift space distortions" (RSD) when mapping the angular position of galaxies and their redshifts into a three-dimensional space. RSD were shown to contain information about the theory of gravity and the matter content of the Universe, and their measurement were first used to constrain cosmology at the turnover of this century \cite{peacock01}. More recently, it has also been suggested to combine RSD with measurements of the Cosmic Microwave Background to measure bulk flows and the Mach number of the universe \cite{atrio_mach}. In addition, and due to general relativistic effects, galaxy redshifts were also found to be sensitive to the difference of potential fields at the emitter's and observer's position, and to the evolution of gravitational potential wells crossed by light photons in their way to the observer (the so-called Integrated Sachs-Wolfe effect, ISW) \cite{sachs_wolfe_67}. The gravitational potential well in clusters of galaxies has only been measured recently through galaxy redshifts \cite{Wotjtak_zs11,Zhao_zs13,kaiser_GCzs_2013,jimeno_Tom_14}, and there is growing interest on all subtle general relativistic effects measurable with redshifts \cite{mcdonald_zphi09,yooetal10,challinor_lewis_GRcls11,BonvinDurrer11,yooetal12,Bonvinetal14,Cai_redshifts_2016}. With the advent of a new generation of galaxy surveys, which are bound to produce redshift measurements massively (either through spectra or multi-band photometry), we propose to study the cosmological redshift as a {\em field}. We will show the cosmological information that redshift anisotropy maps encode, discuss its advantages over traditional clustering statistics, and explore its sensitivity to the physics driving the evolution of the universe.
\\
%
\begin{figure}
\centering
\includegraphics[width=17.cm]{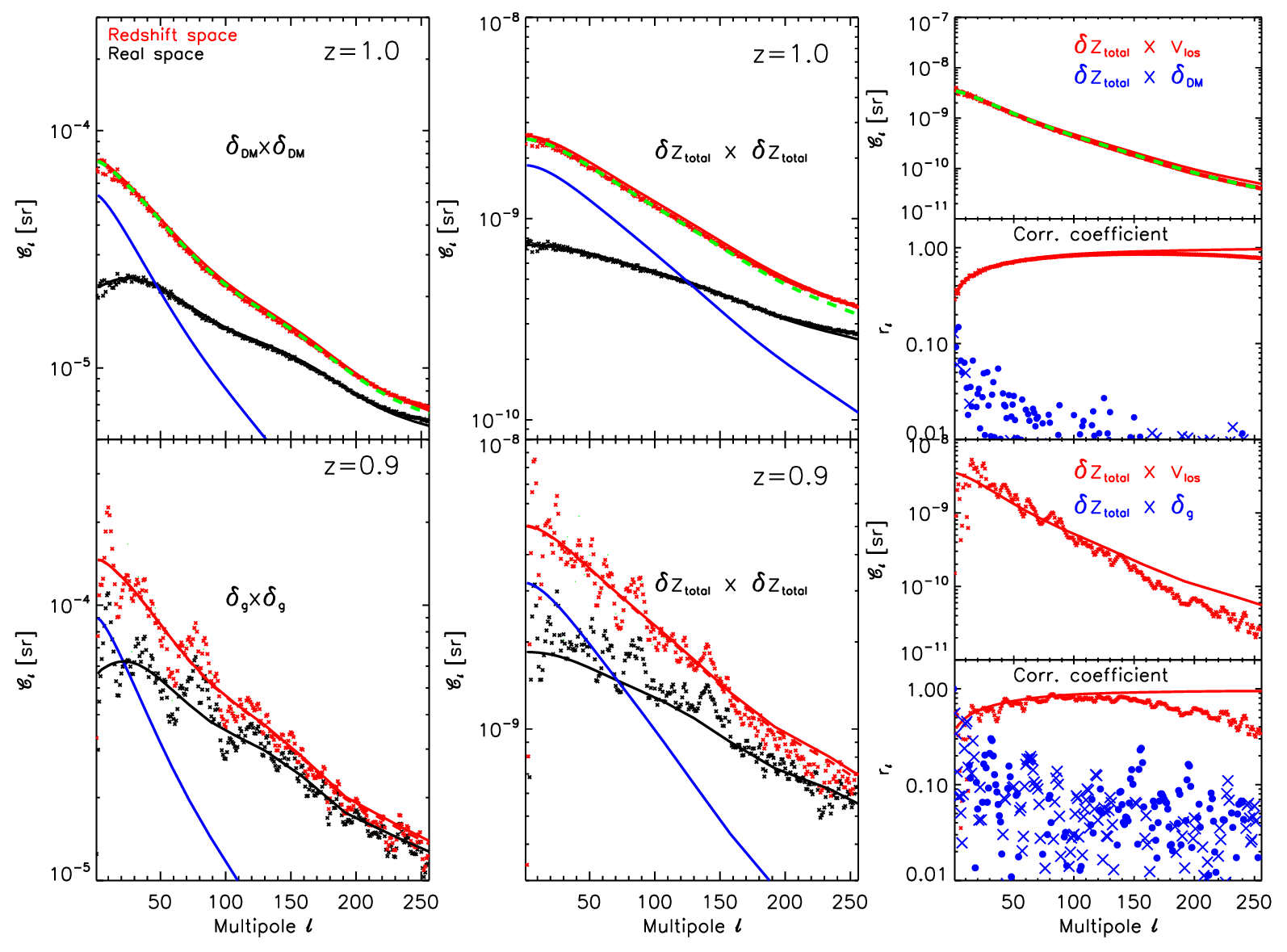}
\caption[fig:F1]{Comparison of linear theory predictions (solid lines) to simulation outputs of the angular power spectra from density contrast (left panels) and from $\dz$ (middle panels). Small crosses refer to the output of N-body simulations, either the average of 100 COLA dark matter particle lightcones (top row), or a galaxy mock from MXXL (bottom row). Results in real and redshift space are given in black and red colors, respectively. Dashed, green lines provide redshift space predictions after considering some radial, thermal, Gaussian motion of particles at the $\sim 450$\,km\,s$^{-1}$ level. The blue, solid lines display the power due to terms containing velocities ($2b_g\,C_{\ell}^{\delta,\,{\rm vlos}}+C_{\ell}^{{\rm vlos},\,{\rm vlos}}$). Right panels display the cross-correlation between $\dz$ and the projected radial peculiar velocity (top sub-panels) and the associated correlation coefficient (bottom sub-panels).}
\label{fig:F1}
\end{figure}

{\bf Methodology.}  Let us consider the angular redshift fluctuations/anisotropies of the redshift field weighted by the galaxy density, $\bar{z} + \dz \equiv \sum_j z_j W_j /\sum_j W_j$, built upon a galaxy sample around the observer's choice for a central redshift $z_{\rm obs}$ and a Gaussian window/weight $W_j = W(z_j; \sigma_z) \equiv \exp{\{-(z_{\rm obs}-z_j)^2/(2\sigma_z^2)\}}$ for the $j$th galaxy falling on pixel $\vn$, and $\bar{z}\equiv\sum_j W_j z_j / \sum_j W_j$. $\dz$ is weighted by the galaxy density in redshift space, and in real space can be written in terms of the underlying matter density and peculiar velocity fields as

\begin{equation}
{\bar z} + \dz = \frac{\int dr \,r^2 \bar{n}(r) \bigl(1+b_g\delta_{\rm m}(r,\vn) \bigr) \bigl( z_H +z_{{\rm vlos}}+z_{\phi} \bigr) W(z_H+z_{{\rm vlos}}+z_{\phi};\,\sigma_z ) }
{\int dr\, r^2 \bar{n}(r)\bigl(1+b_g\delta_{\rm m}(r,\vn)\bigr)W(z_H+z_{{\rm vlos}}+z_{\phi};\,\sigma_z) }.
\label{eq:dz1}
\end{equation}

In this equation, $z_H(r)$ refers to the redshift induced by the Hubble flow and is only a function of the comoving distance $r$ in an isotropic universe; $z_{{\rm vlos}}=(1+z_H)\vvel  (r,\vn)\cdot\vn/c$ is the (position dependent) redshift/blueshift induced by the proper peculiar velocity of the galaxy $\vvel(r,\vn)$ (at linear order in this quantity); $z_{\phi}(r,\vn)$ accounts for redshift fluctuations of gravitational origin and, as will be shown, it provides a negligible contribution. The average number of galaxies at redshift $z_H(r)$ is given by $\bar{n}(r)$ and its bias w.r.t. the total matter distribution is assumed to be constant and equal to $b_g$ for the narrow redshift shells under consideration. The total matter density contrast is given by $\delta_{\rm m}(r,\vn)$, and the volume element of the integrals assume a flat universe. The monopole of the measured redshifts is given by ${\bar z}$.
Under the assumption that perturbations are small ($\delta_{\rm m}, \vvel\cdot\vn/c, z_{\phi} \ll 1$; in particular $z_{{\rm vlos}},z_{\phi}\ll \sigma_z$), one can rewrite Eq.~\ref{eq:dz1} to first order in perturbation theory as
\begin{equation}
{\bar z} +  \dz = \Fcal [z_H] + \Fcal [b_g\delta_{\rm m}\, (z_H-\Fcal[z_H])] +   
  \Fcal \biggl[\biggl(z_{\phi}+\frac{\vvel\cdot\vn}{c}(1+z_H)\biggr) 
     \biggl(1-\frac{d\log{W}}{dz}(z_H - \Fcal[z_H] ) \biggr)
     \biggr] + \,\,{\cal O}(2^{\rm nd}).
      \label{eq:dz2}
\end{equation}
The functional $\Fcal$ applies on a $r$/$z_H$-dependent function $Y(r)/Y(z_H)$ via the normalised integral
\begin{equation}
\label{eq:Fcal1}
\Fcal [Y]  = \frac{\int dr \,r^2 \bar{n}(r) W(z_H;\,\sigma_z) Y(r ) }{\int dr \,r^2 \bar{n}(r) W(z_H;\,\sigma_z) }=  \frac{1} {\cal N} \,\int dr \,r^2 \bar{n}(r) W(z_H;\,\sigma_z) Y(r ).
\end{equation}
In Eq.~\ref{eq:dz2} we have written $z_{\rm vlos}$ explicitly in terms of the peculiar velocity and the Hubble flow redshift.
The logarithmic derivatives of the Gaussian weights ($d\log W/dz$) in that equation are evaluated at $z_{\rm obs}-z_H$. Note that, according to Eq.~\ref{eq:dz2} the redshift monopole ${\bar z}$ equals $\Fcal [z_H]$ (and lies very close to $z_{\rm obs}$ for narrow redshift shells, $\sigma_z<0.03$). It can also be seen that $\Fcal[1-d\log{W}/dz(z_H-\Fcal[z_H])]$ approaches zero for $z_{\rm obs} > 0.1$ and $\sigma_z\lesssim 3\times 10^{-2}$, and this suppresses further the contribution of the integrated Sachs Wolfe effect (ISW) to $z_{\phi}$. In general, the amplitude of $z_{\phi}$ will be typically a couple of order of magnitudes smaller than $z_{\rm vlos}$ and will be ignored hereafter. The above implies that redshift fluctuations that remain roughly constant within the width of the Gaussian window will be severely suppressed, i.e. the $\dz$ field is sensitive to gradients in the abundance of sources within the Gaussian shell. We note we will adopt a Newtonian description, leaving relativistic corrections for future work.

We can now write expressions for $\dz$ and its angular power spectrum $C_{\ell}^{\delta z,\,\delta z}$ in terms of the projected density and peculiar velocity fields in linear perturbation theory (see, e.g., \cite{chm09}). We find that $C_{\ell}^{\delta z,\,\delta z}=b_g^2C_{\ell}^{\delta,\,\delta}+2b_gC_{\ell}^{\delta,\,{\rm vlos}}+C_{\ell}^{{\rm vlos},\,{\rm vlos}}$ where $C_{\ell}^{\alpha,\,\beta}=(2/\pi)\int dk\,k^2 P_{\rm m}(k) \Delta_{\ell}^{\alpha}(k) \Delta_{\ell}^{\beta}(k)$, with $P_{\rm m}(k)$ the linear matter power spectrum extrapolated at present, $\alpha,\beta=\delta_{\rm m}, {\rm vlos}$, $\Delta_{\ell}^{\delta_{\rm m}} = \int dr\,r^2 {\bar n}(r)W(z_H;\,\sigma_z) D_{\delta_{\rm m}}\, (z_H-\Fcal[z_H])\,j_{\ell}(kr)$ and $\Delta_{\ell}^{{\rm vlos}} = \int dr\,r^2 {\bar n}(r)W(z_H;\,\sigma_z) (1+z_H)H(z_H) dD_{\delta_{\rm m}}/dz\,(1-d\log{W}/dz(z_H-\Fcal[z_H]))\,j_{\ell}'(kr)/k$, with $j_{\ell}'(x)\equiv dj_{\ell}(x)/dx=[ \ell\,j_{\ell-1}(x)-(\ell+1)j_{\ell+1}(x)]/(2\ell+1)$. This means that the $\dz$ field is a priori sensitive to the linear matter density contrast growth factor via $b_g\,\sigma_8\,D_{\delta_{\rm m}}$ and the velocity growth factor $(1+z_H)E(z) dD_{\delta_{\rm m}}/dz = f \sigma_8\,D_{\delta_{\rm m}} E$, with $\sigma_8$ the normalization of (linear) density perturbations extrapolated to the present epoch, and $E(z)\equiv H(z)/H_0$ the Hubble parameter normalized to the present epoch ($H(z=0)\equiv H_0$). This dependence on the galaxy bias induces {\it a priori} a dependence on $f_{\rm NL}$, the local non-Gaussian parameter \citep{Afshordi_fNL,Matarrese_fNL_08}. Since this analysis is based upon discrete objects (galaxies and/or quasars), there is also a shot noise contribution giving rise to constant $C_{\ell}$'s, which can be best estimated via Poissonian simulations of the galaxy/quasar distribution.
\\
\begin{figure}
\centering
\includegraphics[width=18.cm]{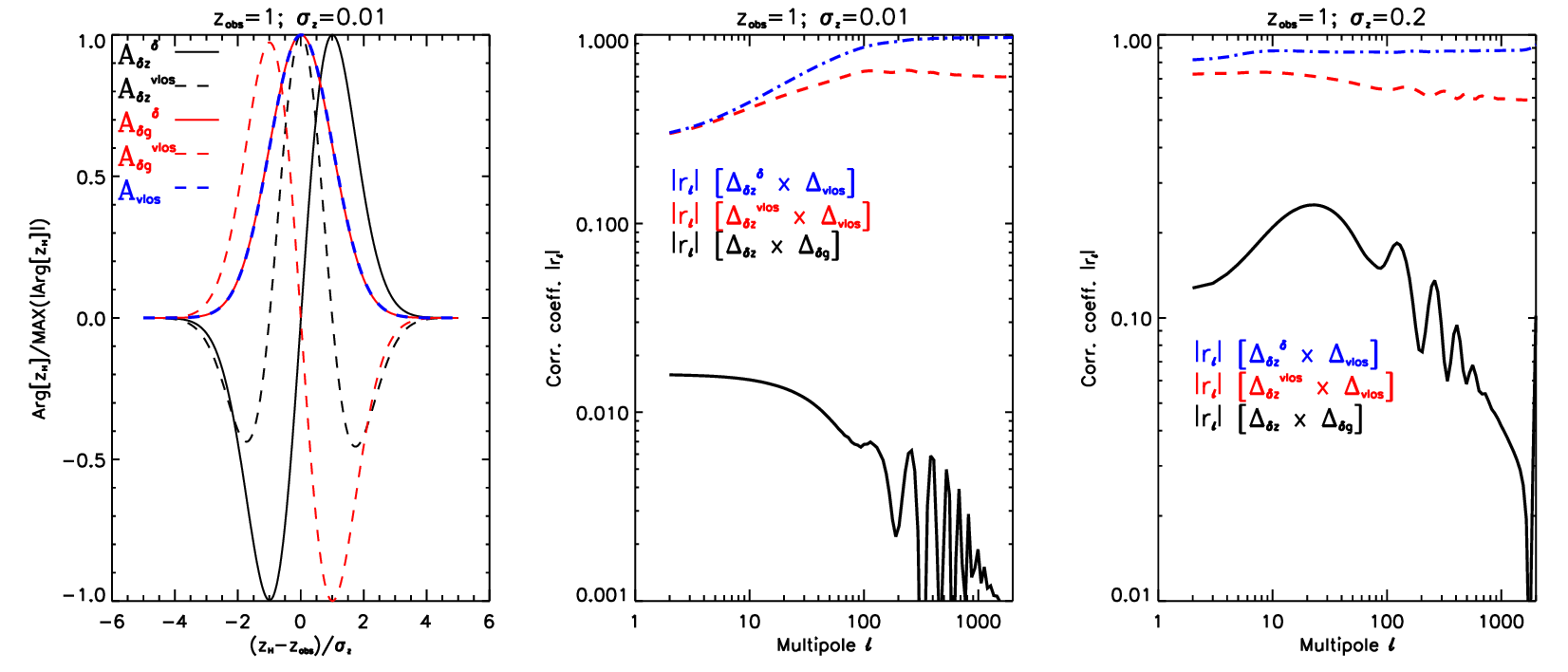}
\caption[fig:F2]{ {\it Left panel:} Radial weight functions (normalized by their maximum absolute value) applied to the terms containing the 3D matter density contrast ($\delta_m$, solid lines) and the radial peculiar velocity ($\vvel\cdot\vn$, dashed lines) in the $\dz$ and $\delta_g (\vn)$ definitions ($z_{\rm obs}=1,\,\sigma_z=0.01$). We also show (in blue color) the radial weight for the radial peculiar velocity projected under the Gaussian window. As an example, according to Eq.~\ref{eq:dz2} the weight function for the $\delta_m$ term contributing to $\dz$ is $W(z_H;\,\sigma_z) (z_H-\Fcal[z_H])$. See text for further details.
{\it Middle panel:} Cross-correlation coefficient of the radial peculiar velocity projected under the Gaussian shell and the $\delta_m$ term of $\dz$ (blue dot-dashed line), the same projected radial velocity term and the $\vvel\cdot\vn$ term of $\dz$ (red dashed line), and the $\dz$ and $\delta_g (\vn)$ fields (black line).  This applies for our reference case $z_{\rm obs}=1 $ and $\sigma_z=0.01$.  {\it Right panel:} Same as middle panel, but for $z_{\rm obs}=1$ and $\sigma_z=0.2$. }
\label{fig:F2}
\end{figure}

{\bf Results.}  We now compare our analytical expressions for $\dz$ with the results of cosmological simulations. Specifically, we compute $C_{\ell}^{\alpha,\,\beta}$ from the outputs of 100 COLA \cite{colaSIMs} simulations presented in \cite{chaves_bao} and a galaxy sample extracted from the MXXL simulation (\cite{Angulo_MXXL}). Unless otherwise stated, in what follows we shall adopt $\sigma_z=0.01$ as our default shell width. For narrow shells and/or low number of galaxies, and in order to avoid instabilities in the denominator, we re-define the $\dz$ estimator as $\dz= \sum_{j\in \vn} W_j (z_j-\bar{z}) / \langle \sum_{j\in \vn} W_j\rangle$, with $\langle ... \rangle$ denoting angular averages in the sky footprint. The same limit for linear theory outlined above applies in this case.  We compare the angular power spectrum, of the projected density contrast and $\dz$ fields (under the same Gaussian window) in the left and middle panels of Fig.~\ref{fig:F1}, respectively. To highlight the impact of peculiar velocities, black color displays the case where redshifts are computed using only positions (real space), whereas red color also considers the contribution of peculiar motions (redshift space). In the top row, crosses refer to the average output obtained from our 100 COLA simulations at $z_{\rm obs}=1$, whereas in bottom panels they correspond to the MXXL results for galaxies with stellar mass above $10^{10}\,h^{-1}$\,M$_{\odot}$ and bias $b_g\simeq 1.3$ ($z_{\rm obs}=0.9$ in this case). Linear theory predictions in real and redshift space are displayed as black and red solid lines, respectively, and the power containing radial peculiar velocity terms is given by the blue solid lines. The green, dashed lines provide theoretical predictions that consider a Gaussian, thermal motion of particles of rms of about $\sim 450$\,km\,s$^{-1}$ along the line of sight. This few-percent correction accounts for deviations due to non-linear evolution and are unnecessary for wider redshift shells ($\sigma_z \gtrsim 0.03$). In our simulations, shot noise lies at a negligible level.

After introducing this correction, the agreement of our theoretical predictions with the average COLA lightcones is typically better than 1\,\% up to $\ell\sim 80$--$100$ in redshift space, and it extends up to $\ell\sim 120$ in real space. On smaller scales, more non-linear effects start to become visible. We see in the middle bottom panel that these non-linearities are more important for the MXXL galaxies, introducing a visible power deficit at $\ell\gtrsim 150$ in redshift space. 

The right panels highlight the correlation existing between the $\dz$ field and the projected radial peculiar velocity field under the same Gaussian redshift window. In both rows, the top sub-panels compare the theoretical prediction for this cross-correlation with the simulations output, whereas the bottom sub-panels provide the correlation coefficient. Predictions from linear theory agree with simulation results up to $\ell\sim 100$--$150$ where non-linear effects kick in. Interestingly, the $\dz$ field is constructed to be almost uncorrelated with the projected density field in redshift space. This is shown in the bottom sub-panels by the blue symbols: the $\dz\times (\delta_m,\,\delta_g)$ cross correlation coefficient multipoles oscillate around zero (crosses and filled circles denote positive and negative values, respectively). 
This virtually null cross-correlation between $\dz$ and the 2D projected density contrast has implications when combining these two observables to constrain cosmology, as we address in the next section. 

%
\begin{figure}
\centering
\includegraphics[width=18.cm]{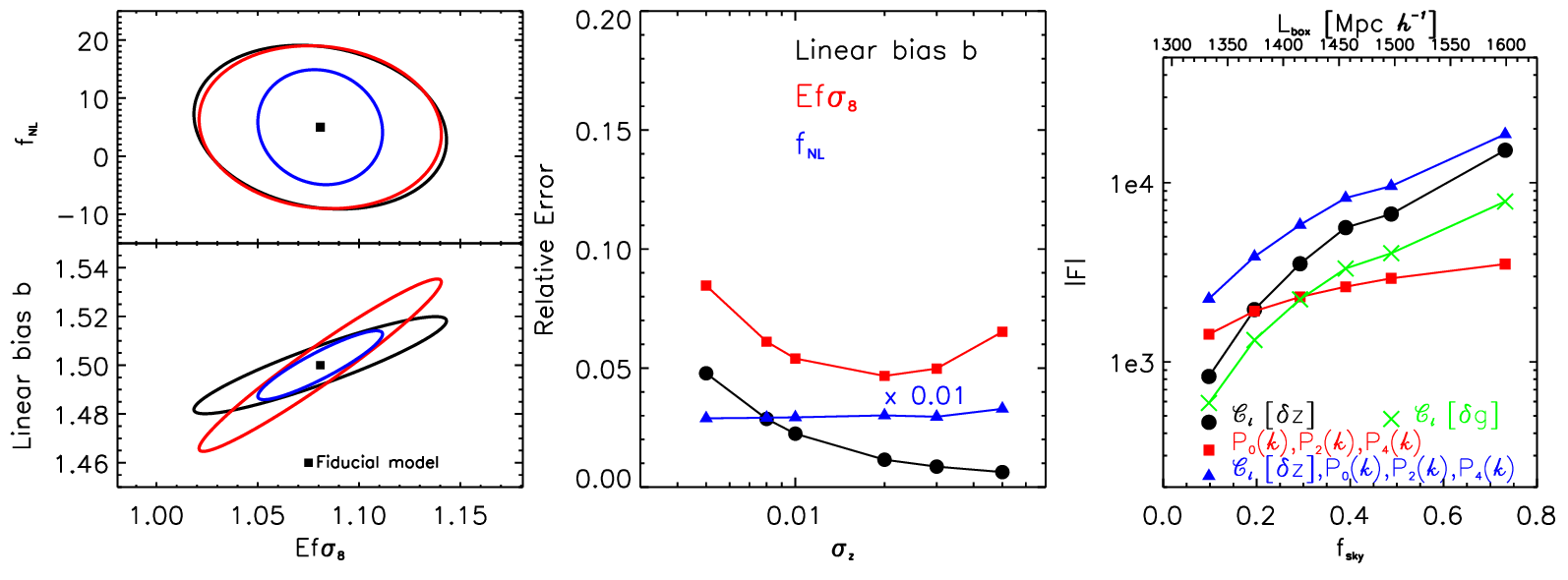}
\caption[fig:F3]{{\it Left panel:} Sensitivity of 2D galaxy density contrast $\delta_g (\vn)$ (black contours) and angular redshift fluctuations $\dz$ (red contours) on the Gaussian linear bias $b_g$, the velocity growth factor $Ef\sigma_8$ and the local non-Gaussian parameter $f_{\rm NL}$. Blue contours provide precision forecasts for the joint analysis of $\delta_g (\vn)$ and $\dz$ after accounting for their covariance. The squares denote our fiducial model. 
{\it Middle panel:} Dependence of the marginalized relative errors in the three target parameters using angular redshift fluctuations with respect to the redshift window width $\sigma_z$. {\it Right panel:} Fisher matrix information versus fraction of sky covered ($f_{\rm sky}$) as provided, for the same cosmological volume, by $\delta z$'s (black, filled circles), standard $P_0(k)$, $P_2(k)$, and $P_4(k)$ estimators (red, filled squares), and the combination of the both (blue, filled triangles). For completeness the Fisher matrix information for $\delta_g (\vn)$ is displayed by green crosses. }
\label{fig:F3}
\end{figure}

This lack of cross-correlation can be intuitively understood by recalling that $\dz$ is sensitive to density gradients within the Gaussian shell, whereas the $\delta_g (\vn)$ field is, by definition, sensitive to the radial (redshift) average of sources under the same shell. This is more quantitatively addressed in Fig.~\ref{fig:F2}. In the left panel we display the (normalized) radial weights applied to the terms involving the 3D matter density contrast ($\delta_m$) and the projected radial velocity ($\vvel\cdot\vn$) employed for estimating angular projections $\dz$ (in black) and the $\delta_g (\vn)$ (in red). 
Radial weights applying to terms containing $\delta_m$ are displayed by solid lines, while those applying to terms containing radial peculiar velocities are given by the dashed lines. These terms do not include the spherical Bessel functions ($j_{\ell}$) nor their derivatives ($j_{\ell}'$), which modulate the terms proportional to $\delta_m$ and $\vvel\cdot\vn$, respectively. This is relevant since these radial integrals are within an integral over $k$-wavemodes: due to the nature of the spherical Bessel functions, the contribution to the $k$-integrals is dominant close to their maximum (occurring at $x\sim \ell$). For a fixed multipole $\ell$, the $k$ integral will be centered upon $k_c\simeq \ell/r(z_{\rm obs})$, so the $j_{\ell}$ function will be positive around its maximum in the full radial integration range under the Gaussian shell shown in the left panel of Fig.~\ref{fig:F2}. However, the derivative $j_{\ell}'$ will flip sign beyond the $j_{\ell}$ maximum (i.e., it will become negative for $z_H-z_{\rm obs}>0$). This means that, in practice, dashed lines in the left panel of Fig.~\ref{fig:F2} will flip sign for $z_H-z_{\rm obs}>0$ when accounting for the $j_{\ell}'$ modulation in the $k$-integral. By taking this into account, and looking at the radial weight functions in Fig.~\ref{fig:F2}, one can infer that projected radial peculiar velocity will be highly correlated to $\dz$, while the 2D $\delta_g (\vn)$ field and $\dz$ will be practically uncorrelated. This is shown in the middle panel of Fig.~\ref{fig:F2}, where the correlation coefficients (defined as $r_{\ell}[A,\,B]\equiv \langle A_{\ell}\,B_{\ell} \rangle / \sqrt{\langle A_{\ell}^2\rangle \langle B_{\ell}^2 \rangle}$) for different pairs of terms are explicitly computed. We remark that this is the case for narrow widths ($z_{\rm obs}=1$ and $\sigma_z=0.01$), being the situation significantly different for wider shells (see right panel in this figure, corresponding to $z_{\rm obs}=1$ and $\sigma_z=0.2$). 
\\

{\bf Discussion and conclusions.}  In Fig.~\ref{fig:F3} we show that the $\dz$ field is sensitive to peculiar velocities and the growth of structure. After considering a single, full sky, redshift shell centered upon $z_{\rm obs}=1$,  with $\sigma_z=0.01$, and a galaxy population of $b_g=1.5$ and arbitrarily high density, we compute Fisher matrix forecasts under a cosmology compatible with the second data release from \planck\,  \cite{planck_parameters_15}, with $f_{\rm NL}=5$. The left panel in Fig.~\ref{fig:F3} shows marginalized 2-D constraints (at 1-$\sigma$) for the three parameters under consideration, namely $f_{\rm NL}$, Gaussian linear bias $b_G$, and the velocity growth factor $Ef\sigma_8$. Black, red, and blue contours refer to constraints set from density contrast alone, $\dz$ alone, and both observables combined, respectively. Since density contrast is uncorrelated to $\dz$ the combination of the two fields improves significantly the precision on the parameters: the Figure of Merit (FoM) values in the $Ef\sigma_8$ vs $b_G$ for density contrast, $\dz$ and the two probes combined are $3,200$, $3,500$ and $9,400$. The corresponding FoM figures in the $Ef\sigma_8$ vs $f_{\rm NL}$ space are $2.12$, $2.04$ and $5.84$. When varying the width of the Gaussian redshift window, we find that values in $\sigma_z\in [0.01,0.03]$ are most sensitive to $Ef\sigma_8$, while for the Gaussian bias and $f_{\rm NL}$  values of $\sigma_z > 0.02$ are preferred (see marginalized relative errors in each of the parameters  in the middle panel of Fig.~\ref{fig:F3}). We next compare the Fisher information associated to the los velocity amplitude for both the $\dz$ field and the standard clustering $P_0(k)$, $P_2(k)$, and $P_4(k)$ power spectrum multipoles. For that, we consider different fractions of sky coverage $f_{\rm sky}$, and impose $L_{{\rm box}}^3 = 4\pi f_{\rm sky}/3\times(r_{z_{\rm max}}^3-r_{z_{\rm min}}^3)$, with $L_{{\rm box}}$ the box size in which the power spectrum momenta $P_i(k)$ are estimated, and $z_{\rm min}, z_{\rm max}=0.8,\,1.2$ the minimum and maximum redshifts in the analysis.
We also assume that the $\delta_{\vk}$ modes are the same in both configurations. We choose $\sigma_z=0.01$ as both the redshift width of the shells and the separation between contiguous shells. For the galaxy population, we adopt the Model 3 from \cite{Pozzetti16} describing the H$_{\alpha}$ emitter galaxy population at $z\sim 1$. We ignore non-linear effects and adopt $k_{\rm max}=0.15\,h$\,Mpc$^{-1}$. The right panel in Fig.~\ref{fig:F3} shows the Fisher information content versus $f_{\rm sky}$ obtained from the $\dz$ field (black, filled circles), 
 the $P_i(k)$'s (red, filled squares), and the combination of the two sets of observables (blue, filled triangles). We account for all covariances among the different $\dz$ fields from different shells, and the (approximate) covariance between the latter and the $P_i(k)$'s, and show that the $\dz$ field contributes with some, additional information, particularly for higher values of $f_{\rm sky}$, mostly due to better sampling of low $k$ power. More detailed computations are deferred for future work.

The $\dz$ field thus allows for a simple, tomographic, direct test of observables (sky coordinates and redshifts) with theoretical predictions for any cosmological setup. We have presented here predictions for first order in linear perturbation theory (LPT) in the $\Lambda$CDM scenario, although there is obvious room for higher order LPT corrections (including general relativistic effects, and/or modified gravity theories). While constraints on cosmological parameters can already be obtained via $\dz$ from spectroscopic surveys like BOSS \citep{boss_dz}, it is also possible to make predictions for spectro-photometric surveys having photo-$z$ precision at the $\sim 1$~\% level or better (like J-PAS). The $\dz$ field may suffer in general for systematics (particularly if redshifts are estimated photometrically). It turns out, however, that angular redshift fluctuations are particularly insensitive to systematics affecting the number of galaxies. 
Indeed, from Eq.~{\ref{eq:dz1}} it is trivial to prove that, given an observed number density of galaxies with multiplicative ($\gamma$) and additive ($\epsilon$) biases ($n^{\rm obs}(\vrv)= \gamma \bar{n} ( 1 + \delta_g (\vrv)+ \epsilon)$), then the observed redshift angular fluctuation field will be $(\delta z)^{\rm obs} (\vn) \simeq \dz + {\cal F}[\epsilon(z_H - {\cal F}[z_H])]$ to first order in the biases. This means that unless the additive bias $\epsilon$ shows a strong redshift dependent throughout the redshift shell, its impact on the observed $\dz$ field should be negligible.

The $\dz$ field can also be a useful observable in cross-correlation studies with maps of the Cosmic Microwave Background (CMB). Given its correlated character to line-of-sight velocities, it constitutes an ideal probe for the kinetic Sunyaev-Zeldovich effect (kSZ) at any redshift \cite{Chaves-Montero_kSZ}. At the same time, the angular redshift fluctuations probe the large scale potential wells, just as angular density fluctuations, and hence can be used to unveil the ISW. For large redshift shell widths, however, redshift and angular density fluctuations become correlated, and thus the increase of sensitivity to the ISW over the standard CMB intensity -- density cross correlation test is very modest.

We have thus demonstrated that density weighted angular redshift fluctuations provide {\em direct} evidence for the distribution of inhomogeneities of matter and peculiar velocities at any cosmological epoch. The interpretation of this observable does not require to convert observed redshifts into distances under any given fiducial cosmological background, and thus is ideal for testing deviations from $\Lambda$CDM. Here we have provided linear theory theoretical predictions for 2-point statistics of this $\dz$ field, and current work is proving $\dz$ to be a competitive cosmological observable in large scale surveys, complementary to other observables used in standard cosmological analyses \cite{boss_dz}. Further studies quantifying the power of this measurable in higher order statistics,  in mildly non-linear scales, in the context of General Relativity, or in  cross-correlation studies to other cosmological observables, are currently ongoing.

\vspace{0.2cm}
\section*{Acknowledgments} We acknowledge useful discussions with G. Aric\`{o}, G. Hurier, E. Komatsu, and J. Miralda-Escud\'e. The authors acknowledge support from the Spanish Ministry of Science, Innovation, and Universities through the projects AYA2015-66211-C2-2 and PGC2018-097585-B-C21, and the Marie Curie CIG9-GA-2011-294183. R.E.A. acknowledges support from the European Research Council through grant number ERC-StG/716151. Argonne National Laboratory’s work was supported by the U.S. Department of Energy, Office of Science, Office of Nuclear Physics, under contract DE-AC02-06CH11357.  This work has made used of CEFCA's Scientific High Performance Computing system which has been funded by the Governments  of  Spain  and  Arag\'on  through the  {\it Fondo  de  Inversiones  de  Teruel}, and the Spanish Ministry of Economy and Competitiveness (MINECO-FEDER, grant AYA2012-30789).
We also acknowledge the use of the {\tt HEALPix}~\footnote{URL: {\tt http://healpix.sourceforge.net}} software \citep{Gorski2005}.
\section*{Data availability}
Data available on request.
\bibliographystyle{mnras}
\bibliography{refs1}

\label{lastpage}
\end{document}